\documentclass{article}
\usepackage{spconf,amsmath,amssymb,graphicx,multirow,booktabs}
\usepackage{microtype}
\usepackage{pifont}
\usepackage[hidelinks]{hyperref}
\usepackage{cite}

\usepackage[flushleft]{threeparttable} 

\title{Contrastive Learning-based Audio to Lyrics Alignment for Multiple Languages}
\name{
Simon Durand\textsuperscript{1}\textsuperscript{*}, Daniel Stoller\textsuperscript{1}\textsuperscript{*}, Sebastian Ewert\textsuperscript{1}
}
\address{
\textsuperscript{1}Spotify
\textsuperscript{*}Equal contribution
}

\begin{document}
\ninept
\maketitle
\begin{abstract}
Lyrics alignment gained considerable attention in recent years. State-of-the-art systems either re-use established speech recognition toolkits, or design end-to-end solutions involving a Connectionist Temporal Classification (CTC) loss. However, both approaches suffer from specific weaknesses: toolkits are known for their complexity, and CTC systems use a loss designed for transcription which can limit alignment accuracy. In this paper, we use instead a contrastive learning procedure that derives cross-modal embeddings linking the audio and text domains. This way, we obtain a novel system that is simple to train end-to-end, can make use of weakly annotated training data, jointly learns a powerful text model, and is tailored to alignment. The system is not only the first to yield an average absolute error below 0.2 seconds on the standard Jamendo dataset but it is also robust to other languages, even when trained on English data only. Finally, we release word-level alignments for the JamendoLyrics Multi-Lang dataset.
\end{abstract}
\begin{keywords}
lyric alignment, music information retrieval, audio signal processing, speech recognition, open-source dataset
\end{keywords}

\section{Introduction}
\label{sec:intro}
For many years, the accuracy of systems for the automatic alignment of audio and lyrics content was well below the requirements for practical applications \cite{mesaros2008alignment,fujihara2006sync,kruspe2018application,wang2004lyrically,chang2017lyrics}.
However, since 2019 there has been a resurgence of research and results improved by an order of magnitude. In this context, given an audio recording and textual lyrics for a piece of music, the goal of automatic lyrics alignment is to identify for each lyrical element the corresponding position in the audio. Hereby, lyrical element can refer to paragraphs, lines, words, phonemes or even characters, depending on the need for temporal precision.

Current systems typically either employ toolkits designed for speech recognition or neural network-based end-to-end systems, both with specific advantages and disadvantages. The former represent a working recipe for many complex systems often centered around a hidden Markov model (HMM) or transducer that transfers a lot of past experience from speech recognition over to lyrics processing. This can yield state of the art results~\cite{gupta2020lyric} and tends to maximize the usefulness of small datasets. However, using tightly integrated components as well as task- and language-specific adaptations based on expert knowledge~\cite{gupta2019acoustic,dzhambazov2015modeling,demirel2021lyric,sharma2019lyrics} such as phoneme dictionaries and duration models, such systems are often considered cumbersome to train and to be limiting experimentation and further progress.
Since the release of larger scale noisy datasets~\cite{meseguer2018dali,smith2013damp} and methods to clean them~\cite{meseguer2020cleansing}, many efforts~\cite{watanabe2018espnet,gulati2020conformer} have focused on building conceptually simpler end-to-end systems that enable quick iteration and can make use of larger amounts of data. Many of these systems are trained using the Connectionist Temporal Classification (CTC) loss \cite{graves2006connectionist}, which is a specific instance of the forward-backward (FB) algorithm \cite{rabiner1989tutorial}.

CTC-based training was first employed in \cite{stoller2019lyric} for lyrics alignment and was further explored in \cite{vaglio2020lyric,teytaut2021ctc,jiawen2022lyric}. While CTC enables training neural networks from unaligned lyrics-audio pairs and thus from larger datasets, there are several drawbacks to using CTC for alignment purposes. First, CTC acts as a transducer by introducing an $\epsilon$ symbol (see \cite{graves2006connectionist} for details), which means 'no output' for an audio frame. It was originally used to enable HMM-less decoding of neural network outputs for transcription, but can lead to ill-defined target paths when used for alignment. Second, CTC assumes the text and audio sequences match one-to-one except for temporal distortion and thus it is not prepared for missing or extra symbols in the annotation - a problem more severe for lyrics than speech due to the prominence of unannotated instrumental sections and general annotation errors.
Third, CTC at training time is linear in the number of symbols it supports, which means that in practice CTC systems only support characters or phonemes but not (sub-)words - yet these could be detected more robustly. As a consequence, CTC does not support context-dependent phones while the surrounding context within a word affects phonetic variability greatly.

Inspired by embedding-based approaches in neighbouring fields~\cite{wu2022wav2clip,momeni2020keyword,schulze2021phoneme,mithun2019weakly}, we present a novel end-to-end method that represents the audio signal and the lyrics as sequences of embeddings and uses contrastive learning to map them into a joint space without requiring additional information.
This way, we obtain a stable loss function that does not require an $\epsilon$ symbol or a fixed in-memory state space, allowing us to model characters in context with their neighbours, which would otherwise lead to an infeasibly large vocabulary due to combinatorial explosion.
Due to this data-driven, neural text representation, our system is easy to extend to additional languages, and more robust than modelling each character independently.
We only require an unordered bag of words occurring in an audio segment to train the system, opposed to conventional methods that require an ordered, complete word sequence or even symbol timings.
It performs at the level of the state-of-the-art, yielding an average absolute error below 0.2 seconds on the standard JamendoLyrics dataset~\cite{stoller2019lyric} for the first time.
We also release manually annotated word-level alignments for an extension of that dataset featuring three additional languages.
This enables a robust evaluation of multi-language lyrics alignment approaches, whereas prior work in this setting~\cite{vaglio2020lyric,brazier2021lyrics} relied on performance estimates from noisy annotations~\cite{meseguer2018dali}, or required duplicate songs.

In the following section~\ref{sec:similarity}, we will present the core elements of our approach: the text and audio encoders, the similarity matching and training procedure, and the line-based decoding.
Section~\ref{sec:experiments} shows, via an evaluation on the extended JamendoLyrics Multi-Lang dataset, the relative advantages of our main design choices, and how the approach is both competitive with the state-of-the-art and can scale well to additional Latin languages. Finally, we conclude and present future directions in section~\ref{sec:conclusion}.

\section{Similarity model}
\label{sec:similarity}

\begin{figure}[t]
  \centering
  \includegraphics[width=7.5cm]{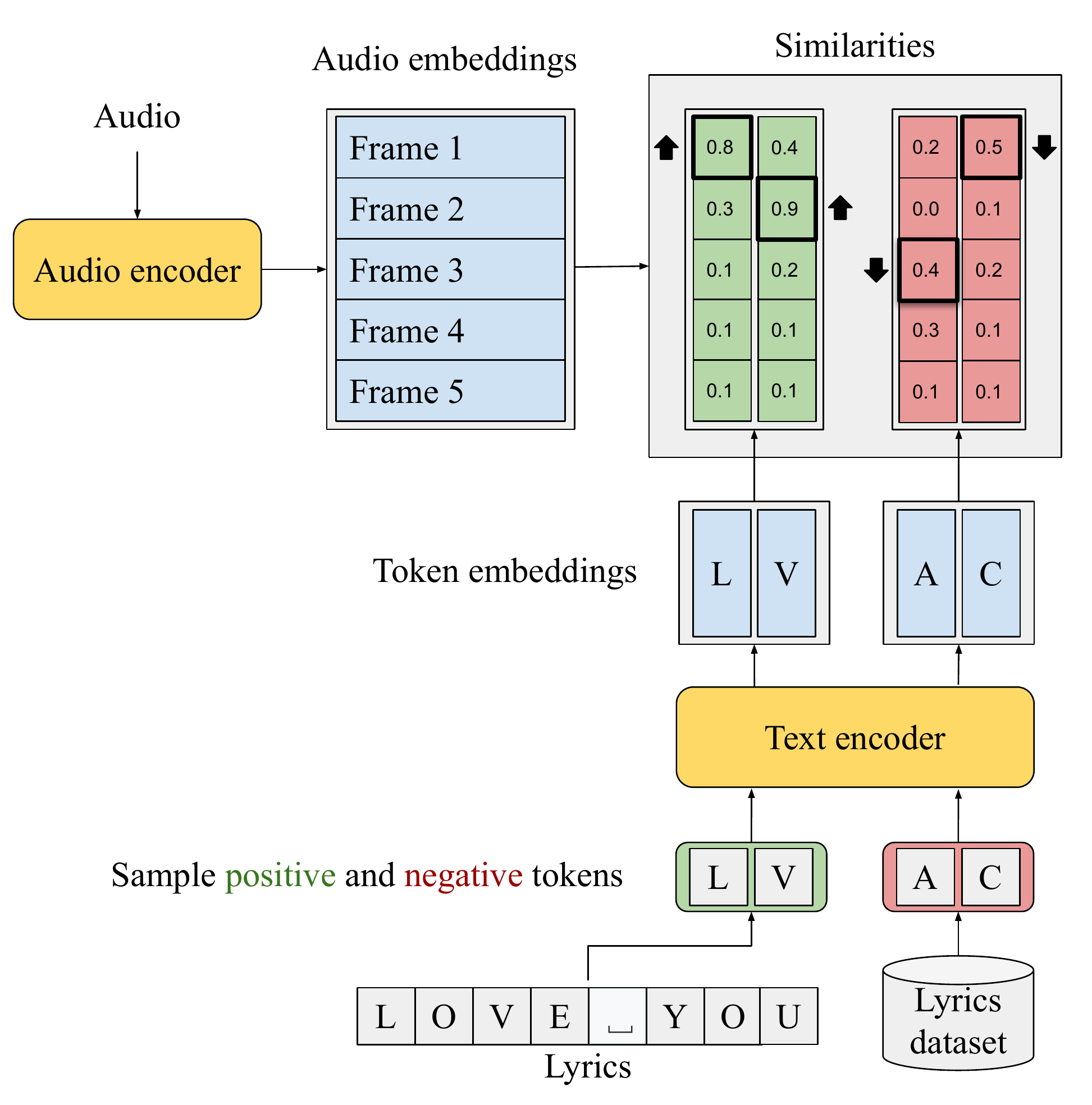}
  \vspace{-0.1cm}
  \caption{\footnotesize Our proposed approach training for an audio-lyrics pair. Positive lyrics tokens are sampled from the lyrics, while negative ones are taken from lyrics of other songs in the dataset. The audio and text encoders produce embeddings for the audio and lyrics tokens, respectively. Finally, for each lyrics token embedding, the maximum similarity to the audio embeddings is maximised (positive examples) or minimized (negative examples).}
 \label{fig:sim_training_overview}
 \vspace{-0.1cm}
\end{figure}

An overview of our proposed similarity model and its training is given in figure~\ref{fig:sim_training_overview}.
In the following, we describe its audio and text encoder components, the contrastive learning procedure for training, and the alignment decoding.
\vspace{-0.2cm}

\subsection{Audio encoder}
\label{ssec:audio_encoder}

The first component of our similarity model is the audio encoder $f_a$, which aims to detect the phonetic content of the singing voice in the audio.
It receives features $\mathbf{X} \in \mathbf{R}^{T \times D}$ that represent the audio content with $D$-dim. feature vectors over time $t \in \{1,T\}$.
In this paper, we use spectrogram representations with 5 seconds duration, magnitudes normalized by $y \rightarrow \log(1+y)$, a sampling rate of 11025 Hz, an FFT size of 512 and a hop size of 256. 
Other representations are also possible, as long as their temporal resolution is high enough for subsequent alignment.
For the encoder itself, we use a residual network with 10 residual convolutional blocks (RCBs).
Each RCB consists of 2 repetitions of the following sequence of operations: group normalization~\cite{wu2018group}, ReLU activation, and a 2D convolutional layer with a (3x3) kernel and 64 features.
The output of the above is added to the input to the RCB.
Finally, a 1D convolution layer is applied on each time bin with $E$ filters of kernel size $D$ to eliminate the frequency dimension and yield an embedding matrix $\mathbf{A} \in \mathbf{R}^{T \times E}$ with $E=64$ as the embedding size.
This output, with a purposely small receptive field of 930 ms, is supposed to describe the local phonetic content of the singing voice in the audio to enable matching with the lyrics text.

\vspace{-0.2cm}
\subsection{Text encoder}
\label{ssec:text_encoder}

With an estimate of how the singing voice could sound for any part of the given lyrics text, we can match the text parts to the representation from the audio encoder.
This estimation is performed by our text encoder $f_\ell$, which receives the lyrics text as a sequence of symbols $s_1, \ldots, s_N$, where the symbols could correspond to characters, phonemes or other text representations.
To process these symbols, we pass them through a trainable embedding layer that maps each symbol to a different embedding vector, resulting in an embedding matrix $\mathbf{L} \in \mathbf{R}^{N \times E}$.

The resulting embeddings describe the likely phonetic content independently for each symbol.
However, the pronunciation of a symbol usually depends on the neighbouring symbols in the text, especially when using characters as symbols.
Therefore, we extend the text encoder to process the subsequence $(s_{n-C},s_{n-C+1},\ldots,s_{n},\ldots,s_{n+C})$ for each symbol $s_n, n \in \{1,\ldots,N\}$, containing the $C$ previous and following symbols as context information.
A special padding symbol is used for $s_n$ with $n<1$ or $n>N$ to ensure an equal length of each subsequence.

In case we know the language of the input song\footnote{As lyrics are part of the input, this is a reasonable assumption. For instance, there is 100\% accuracy on JamendoLyrics Multi-Lang using~\cite{joulin2016bag}.}, the text encoder could exploit this information to better estimate the  pronunciation.
In this language-conditioned setting, we create a trainable embedding for each language and append it to each text subsequence.
In this paper, we use a simple network with one fully connected layer (three in case of language conditioning) with ReLU activation, and a linear output layer to yield an $E$-dim. embedding for each symbol -- although more complex architectures can be used for further improvements.
Both the text and audio encoder embeddings are L2 normalized to enable cosine similarity comparisons.

\vspace{-0.3cm}
\subsection{Similarity matching and training}
\label{ssec:matching}

The audio and text encoders project the audio and lyrics into a shared embedding space.
We aim to train the model such that these similarities are high for matching audio and text parts and low for non-matching ones.
If we knew the exact position of each symbol in the audio, this could simply be achieved by training a classifier where the correct symbol for each position in the audio has to be identified, and each class is equal to one (unique) symbol from the current lyrics sequence.
However, such strong labels are difficult to obtain with sufficient accuracy, so we consider the case where we only know the start and end of the lyrical lines in the audio.

In this weak label case, we cannot apply such a simple classification objective.
We also do not want to use an auto-regressive sequence-to-sequence model objective~\cite{graves2013speech}, as it incentivizes a strong language model, which we found in our early experiments to be detrimental to alignment if it inhibits the acoustic model.
Instead, we apply contrastive learning, where a positive example $s^+$ is taken from the lyrics for a given audio segment and negative examples $s^-$ are sampled from the distribution $p_s$ over symbols obtained from all lyrics in the dataset that do \emph{not}\footnote{If unannotated symbols are present in the audio and mistakenly used as negatives, our method still works as the majority of examples are correct.} appear in the audio segment.
To estimate whether a symbol $s$ appears somewhere in the audio, we take its maximum similarity over the whole audio features $\mathbf{X}$, $m(\mathbf{X},s) = \max_t f_\ell(s) \cdot f_a(\mathbf{X})_t^T$, and apply the training objective:
\begin{equation}
L = \mathbb{E}_{(\mathbf{X},s^+) \sim p_d} \left[ (m(\mathbf{X},s^+) - 1)^2 + \mathbb{E}_{s^- \sim p_s} m(\mathbf{X},s^-)^2 \right]
\label{eq:contrastive_loss}
\end{equation}
where $p_d$ is the distribution over audio examples and symbols sampled from the corresponding lyrics sequence.
The loss drives the maximum similarity between positive audio-symbol pairs up to 1.
Negative pairs are pushed towards 0 to make such audio and text embeddings orthogonal.
\subsection{Decoding}
\label{ssec:decoding}

After training the model on pairs of audio and text fragments, we perform the alignment by computing a normalised similarity matrix between audio and text sequences: $\mathbf{S} = \frac{1}{2}(\mathbf{A} \cdot \mathbf{L}^T + 1)$, with $\mathbf{S} \in [0,1]^{T \times N}$.
We then decode an alignment from $\mathbf{S}$ by finding a monotonic path maximizing the cumulative similarity score.

With the above decoding, we found our model to mistakenly position the last few words of a lyrical line close to the start of the next one (and similarly, the first few words of a lyrical line at the end of the previous one).
To alleviate this problem, we exploit that lyrics are temporally clustered and constrain the model to output all words from the same line close to each other. 
From the initial alignment obtained with $\mathbf{S}$, we estimate each line interval as starting at $t_s = t_c - \frac{t_d-d}{2}$ and ending at $t_e = t_c + \frac{t_d+d}{2}$. 
Here, $t_c$ represents the center time of the line by taking the estimated start time of its middle token, which we found to be robust to outliers.
$t_d$ is the duration of the lyrical line, estimated by multiplying the number of tokens in the line with an estimated duration $d$ per token (which we empirically set to $0.2$s for characters and $0.4$s for phonemes but could be made token- and song-dependent to improve performance).

\begin{figure}[t]
  \centering
  \includegraphics[width=8.5cm]{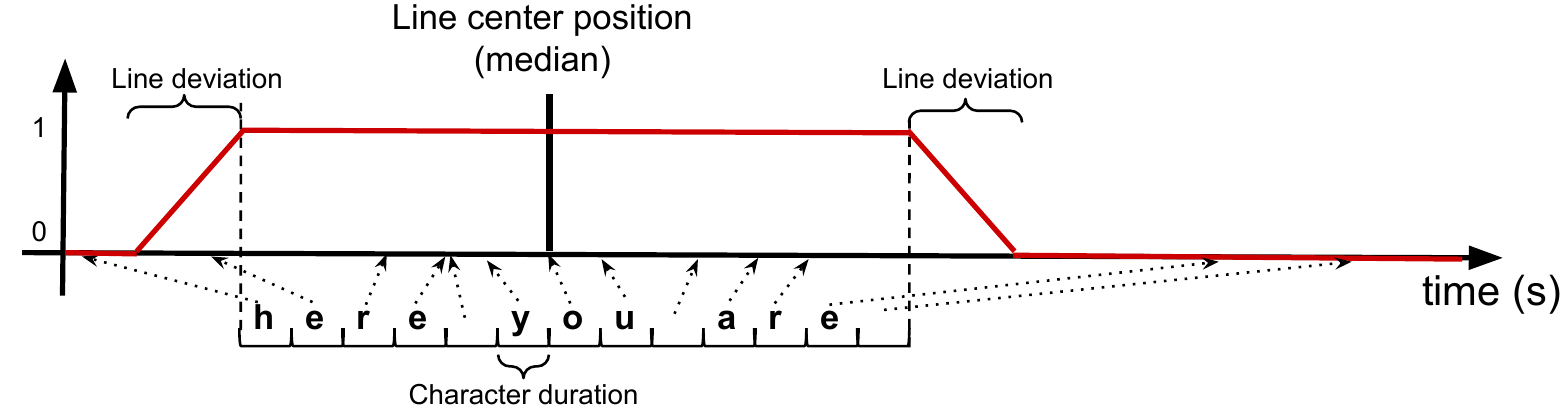}
  \caption{\footnotesize Constructing one row of the line-mask $\mathbf{M}$ for the lyrical line ``Here you are'' (red line), used during line-based decoding. Dashed arrows correspond to the character's start time estimated in the first decoding. The initial estimates for ``h'' in ``here'' and ``re'' in ``are'' are moved close to the other characters in the line by the line-based decoding.}
 \label{fig:sim_mask}
 \vspace{-0.3cm}
\end{figure}

Using that estimated line position, we define a line-mask $\mathbf{M} \in [0,1]^{T \times N}$ for $\mathbf{S}$, with each column of $\mathbf{M}$ constructed as shown in figure~\ref{fig:sim_mask} and constraining each token to be aligned around its estimated line position. Note that $\mathbf{M}$ does not require any additional training or external system, as opposed to using an external vocal activity~\cite{demirel2021lyric} or boundary~\cite{jiawen2022lyric} detection module, and is really fast to obtain.
The final alignment is produced by applying the initial decoding scheme again, but this time to the masked similarity matrix $\mathbf{S}\circ\mathbf{M}$, with $\circ$ as the Hadamard product.
As a result, high values within the line, as seen in figure~\ref{fig:sim_mask}, encourage the model to position all the tokens of a line within the initially estimated line interval $(t_s,t_e)$.
To account for errors in the initial estimation, a linear tolerance window around the line borders is used.
We empirically set the length of this window to $2.5$s, but find performance to be quite robust to this parameter.

\section{Experiments}
\label{sec:experiments}
\vspace{-0.3cm}
\subsection{Methodology}
\label{ssec:method}

We use two standard evaluation metrics, the average absolute error (AAE) in seconds and the percentage of correct onsets according to a tolerance window of 0.3 seconds (PCO), due to their complementary nature:
the AAE, where smaller is better, is sensitive to both very small and especially very large positioning errors, whereas PCO, where higher is better, encourages predicting timestamps that are acceptably close to the ground truth while treating all types of errors the same.
For training, we use a dataset of professional-quality recordings featuring English, Spanish, German and French songs.
Our models are trained for at most 100 epochs with 20,000 iterations each, using an ADAM optimizer with learning rate $0.001$.
An epoch contains approximately 400 hours of data, sampled among a collection of 87,785 songs.
We reserve 2\% of our data for validation and compute the loss after each epoch for early stopping with a patience of $20$ epochs. Our model has 1.2 Million parameters for a modest total size of 4.8MB. All our phoneme experiments are based on the CMU English pronunciation dictionary\footnote{\url{http://www.speech.cs.cmu.edu/cgi-bin/cmudict}}.
For each example, and we use all symbols as positive samples together with 1000 negative samples.

We use the openly available JamendoLyrics Multi-Lang dataset~\cite{demirel2023submission} for evaluation, that features Creative Commons songs and is an extension of the JamendoLyrics dataset~\cite{stoller2019lyric} to German, French, and Spanish (60 additional songs with a high genre diversity) on top of the existing English data. We extend this dataset with manually annotated word-level timings for all songs. We do not evaluate on other commonly used datasets as they are smaller, only support English recordings and might overlap with our training set.

\vspace{-0.2cm}
\subsection{Results and discussion}
\label{ssec:results}

We start by evaluating different configurations of our system, to assess the impact of the loss, contextual embeddings, the choice of text representation, and the use of line-based decoding. We then compare our approach with the current state-of-the-art and conclude by investigating the multilingual performance.

\emph{How does CTC compare to our contrastive learning approach?} To evaluate against a CTC approach, we build a CTC-based model using  the audio encoder from our similarity model, leaving the data and experimental setup unchanged. We add a fully connected layer to the audio encoder output to get phonemes probabilities, and use the CTC loss during training. 
As shown in Table~\ref{table:configs-en} (model M1), the performance is similar to the CTC approach from~\cite{stoller2019lyric} shown in Table~\ref{table:sota}. However, its AAE is significantly worse than our baseline similarity model (M3) using the same, phoneme-based configuration. Our proposed text encoder together with contrastive learning then appears to yield better performance than a simple CTC approach.

In the above comparison (between models M1 and M3), no additional prior knowledge is used to constrain the models for further performance improvements.
However, high-performance CTC systems~\cite{vaglio2020lyric, jiawen2022lyric} are indeed incorporating additional information by adding vocal separation, pitch information or external boundary estimation.
Similarly, we also obtain a competitive AAE for the CTC approach (M2) when swapping the mask $\mathbf{M}$ used for line-based decoding with the one obtained from our best similarity model. This suggests CTC approaches rely heavily on such additional constraints, whereas our model is more robust while being comparatively simple.

\emph{What is the impact of a contextual text representation?} We can compare similarity models processing multiple symbols with the text encoder (models M5 and M6) and those processing a single symbol without additional context (models M3 and M4) in Table~\ref{table:configs-en}. We see that processing multiple (M5) rather than single (M3) phonemes leads to significantly stronger results\footnote{We use small values for $C$ as long-range text dependencies are not crucial for determining the sound of each symbol.}. This effect is even stronger for characters, where processing characters independently (M4) leads to poor performance, and using multiple characters (M6) yields an AAE of 0.15s, the lowest reported in the literature so far.
The above results show the benefits of modeling a symbol based on contextual information influencing its pronunciation. We hypothesize this contextual information also leads to less spurious activations by non-vocal sounds. For instance, individual symbols are more easily mistaken with an instrumental section than sequences of symbols. It may also make negative sampling more discriminative. These contextual text embeddings cannot be used with CTC approaches because of a combinatorial explosion when considering all possible sequences of symbols as individual classes, and because of the limited support for text modeling. 
This might be one reason why recent CTC-based works use external information to yield a reduced search space and work well. 
Unlike standard toolkits that can use triphone states, we do not have to deal with a large increase in vocabulary size. Furthermore, our approach learns from the data directly so we do not have to manually model context-dependant text pronounciation with heuristics that are hard to scale to different languages and music styles.

\emph{Is a phoneme representation mandatory?} If the text encoder does not process symbols with context, we see that a phoneme representation (model M3) is important to get a good performance, compared to using a character representation (model M4), as indicated by prior research~\cite{gupta2020lyric, demirel2021lyric, jiawen2022lyric, vaglio2020lyric}. However, when the text encoder processes characters with context, then a phoneme representation (model M5) performs similarly as a character representation (model M6). We hypothesize that contextual character representations could be thought as a form of data-driven and end-to-end phoneme representation.

\emph{What is the effect of a line-based decoding?} When we skip the line-based decoding (M7) of the best character model, we observe a significantly worse AAE. It can indicate that the mask $\mathbf{M}$ indeed helps removing outliers. We note that all configurations worked better using this simple decoding process.

\begin{table}[t]
\centering
\begin{threeparttable}
\begin{tabular}{ccccccc}
 \toprule
 Model & Loss & Token & $C$ & LB-Dec. & AAE & PCO \\
 \midrule
 M1 & CTC & Phon & 0 & \checkmark &0.90 & 86 \\
 M2 & CTC & Phon & 0 & \checkmark \tnote{*} &0.20 & 89\\
 M3 & Sim & Phon & 0& \checkmark& 0.39 & 89 \\
 M4 & Sim & Char & 0& \checkmark & 1.60 & 73 \\
 M5 & Sim & Phon & 1 & \checkmark & 0.16 & \textbf{93}\\
 M6 & Sim & Char & 1 & \checkmark& \textbf{0.15} & 92\\
  M7 & Sim & Char & 1 & \ding{55} & 0.24 & 91 \\

 \bottomrule
\end{tabular}
\begin{tablenotes}
\footnotesize 
\item[*] Uses mask matrix $\mathbf{M}$ from the best similarity model
\end{tablenotes}
\end{threeparttable}
\caption{\footnotesize Different model configurations trained on English data only and evaluated on the JamendoLyrics English dataset, using the CTC or similarity (``Sim'') model, a character (``Char'') or phoneme (``Phon'') representation, and $C$ context symbols as described in section~\ref{ssec:text_encoder}. ``LB-Dec.'' refers to line-based decoding from section~\ref{ssec:decoding}.}
\label{table:configs-en}
\vspace{-0.2cm}
\end{table}

\begin{table}[t]
\centering
\begin{threeparttable}
\begin{tabular}{p{1.50cm}lcc}
 \toprule
 System & Dependencies\tnote{*} & AAE (s) & PCO (\%) \\
 \midrule
 SDE~\cite{stoller2019lyric}& None & 0.82 & 85\\
 VHMRA~\cite{vaglio2020lyric}& SS & 0.37 & 92 \\
 DAD~\cite{demirel2021lyric}& Tr, Pho& 0.31 & 93 \\
 HBE~\cite{jiawen2022lyric}& SS, Pitch, Pho, Boun & 0.23 & \textbf{94} \\
 GYL~\cite{gupta2020lyric}& Gen, Pro, Mult, LM, Pho & 0.22 & \textbf{94} \\
  M5& Pho & 0.16 & 93 \\
 M6& None & \textbf{0.15} & 92 \\
 \bottomrule
\end{tabular}
\begin{tablenotes}
\item[*] \footnotesize SS: Source separation. Pho: Grapheme-to-phoneme model. Pitch, Gen: Uses pitch or genre information during training. Tr: Transcription model. B: Boundary model. M: Multiple alignment models. Pro: Pronunciation model. LM: Language model.
\end{tablenotes}
\end{threeparttable}
\caption{\footnotesize Comparison with published methods on the JamendoLyrics English dataset, and their dependencies. We see that previously, more dependencies increased performance, at the cost of additional complexity.}
\label{table:sota}
\vspace{-0.5cm}
\end{table}

\begin{table}[t]
\centering
\footnotesize
\begin{tabular}{lcp{0.4cm}p{0.4cm}p{0.4cm}p{0.4cm}p{0.4cm}p{0.4cm}}
 \toprule
 & & \multicolumn{2}{c}{All} & EN & ES & DE & FR \\
 \cmidrule(lr){3-4}
 Model & Data & AAE & PCO & AAE & AAE & AAE & AAE \\
 \midrule
 M5 & EN& 1.11 & 59 & 0.16& 1.72&1.22  & 1.36\\
 GYL~\cite{gupta2020lyric}& EN & 
 0.65& 73& 0.22 & 0.97 & 0.62 &0.78\\
 M6 & EN& 0.35 & 89 & \textbf{0.15} & 0.37& 0.41 & 0.47\\
 M6 & All & 0.29 & 91 & 0.39 & 0.22 & 0.26 & 0.28\\
 M6 Lang-Cond. & All & \textbf{0.18} & \textbf{94} & 0.21 & \textbf{0.13} & \textbf{0.16} &\textbf{0.19}\\
 \bottomrule
\end{tabular}
\caption{\footnotesize Evaluation of different systems on the JamendoLyrics Multi-Lang dataset, training on English data only (``EN'') or on English, Spanish, German, and French (``All'').}
\label{table:configs-all}
\vspace{-0.35cm}
\end{table}

\emph{How do similarity models compare to the state-of-the-art?} We compare our results with several published models in Table~\ref{table:sota}. We see both the phoneme and character similarity models surpass the state-of-the-art for the AAE metric, and stay close to the state-of-the-art for the PCO metric, while requiring at the same time less dependencies. 

\emph{How do similarity models perform on multilingual data?} To estimate the multilingual performance of published models trained on English data, we apply the NUS AutoLyricsAlign software\footnote{\url{https://github.com/chitralekha18/AutoLyrixAlign}} on the JamendoLyrics Multi-Lang dataset, and evaluate the alignment\footnote{While there exists a multi-language model~\cite{vaglio2020lyric}, we could not find a way to apply it to this dataset, and we want to avoid evaluation on a dataset with overlapping train and test data and potentially inaccurate annotations.}. We find that the performance on English matches what was published by the authors~\cite{gupta2020lyric}. In Table~\ref{table:configs-all}, we compare this phoneme model against our phoneme (M5) and character (M6) variants, also trained on English data only. We see the phoneme models don't retain their strong English performance on other languages while the character-based variant manages to maintain relatively high performance on non-English data, with an AAE almost halved compared to~\cite{gupta2020lyric}. This highlights the risk of error propagation of phoneme models if we try to extend the scope to additional languages, and that the performance is limited by the performance of the external phoneme representation.

As expected, including non-English training data (for model M6) increases performance on the non-English subsets, but its performance on English decreases. Importantly, adding a language conditioning layer to M6 (last row of Table~\ref{table:configs-all}), as described in section~\ref{ssec:text_encoder}, enables a strong performance on all languages while remaining competitive with the current state-of-the-art on English data -- indicating we successfully leverage knowledge of lyrics language without language-specific modelling effort. The model sets a new benchmark for multilingual lyrics alignment.

\vspace{-0.25cm}
\section{Conclusion}
\label{sec:conclusion}
\vspace{-0.2cm}
In this paper, we presented a novel lightweight system tailored to lyrics alignment featuring an audio and text encoder, and trained using contrastive learning on weakly labelled data. The alignment is performed on a similarity matrix of text and audio embeddings, and the proposed approach reaches the state-of-the-art on English.
We show that processing lyrics characters with a context window in an end-to-end fashion is key to reach the accuracy of a handcrafted phoneme representation, and to generalise to unseen languages.
With the addition of a simple and easily usable language-conditioning, we obtain strong performance across all languages.
We share word-level alignments for the JamendoLyrics Multi-Lang dataset that researchers can extend to a more diverse set of languages, and use to evaluate multi-lingual alignment systems rigorously.
We believe our embedding-based, data-driven approach is key to address low-resource languages by means of transfer and few-shot learning, as the same model architecture can be reused across languages. As our contrastive learning approach can be trained on audio where the lyrics overlap in time (e.g. duets, Ad-Lib) since it does not need information about word order, it would be interesting to explore this direction and adapt the decoding stage to enable new types of alignment.

\vfill\pagebreak

\bibliographystyle{IEEEbib}
\bibliography{strings,refs}

\end{document}